\documentclass[reprint,aps,prl,superscriptaddress]{revtex4-2}
\usepackage{graphicx}
\usepackage{dcolumn}
\usepackage{bm}
\usepackage[usenames]{color}
\usepackage{xcolor}
\usepackage{tabularx}
\usepackage{booktabs}
\usepackage{hyperref}
\usepackage{siunitx}
\usepackage[utf8]{inputenc}
\usepackage[english]{babel}
\hypersetup{colorlinks,linkcolor=red,citecolor=blue,urlcolor=blue,final}
\usepackage{amsmath, amsthm, amscd, amssymb} 
\usepackage[mathlines]{lineno}

\begin{document}

\title{3D trapping of a meta-atom in an intensity minimum}

\author{Bin Lu}
\affiliation{Nanophotonic Systems Laboratory, Department of Mechanical and Process Engineering, ETH Zurich, 8092 Zurich, Switzerland}
\affiliation{Quantum Center, ETH Zurich, 8083 Zurich, Switzerland}

\author{Adeel Afridi}
\affiliation{Nanophotonic Systems Laboratory, Department of Mechanical and Process Engineering, ETH Zurich, 8092 Zurich, Switzerland}
\affiliation{Quantum Center, ETH Zurich, 8083 Zurich, Switzerland}

\author{Nadine Meyer}
\email{nmeyer@ethz.ch}
\affiliation{Nanophotonic Systems Laboratory, Department of Mechanical and Process Engineering, ETH Zurich, 8092 Zurich, Switzerland}
\affiliation{Quantum Center, ETH Zurich, 8083 Zurich, Switzerland}

\author{Romain Quidant}
\email{rquidant@ethz.ch}
\affiliation{Nanophotonic Systems Laboratory, Department of Mechanical and Process Engineering, ETH Zurich, 8092 Zurich, Switzerland}
\affiliation{Quantum Center, ETH Zurich, 8083 Zurich, Switzerland}

\date{\today}
\begin{abstract} 

High–refractive-index particles have recently attracted a growing interest in optical manipulation, enabling to engineer optical forces through electromagnetic Mie resonances. Unlike standard silica particles, which exhibit trap frequencies mainly determined by material density, resonant meta-atoms formed by high–index particles enable qualitatively new trapping behaviors. In this work, we experimentally investigate the trapping of resonant silicon particles in an optical standing wave. A direct comparison of silicon and silica highlights the fundamental differences in their optical force scaling and trapping dynamics. Beyond conventional trapping at intensity-maxima, we demonstrate deterministic and stable three-dimensional trapping of silicon nanoparticles in optical intensity minima, a regime that remains inaccessible for silica particles. Drawing a mesoscopic analogy with blue-detuned atom trapping, our results establish meta-atoms as a versatile approach to further extend the optical manipulation tool box towards accessing novel trapping regimes e.g. in close proximity to a surface.
\end{abstract}

\maketitle

Since its discovery in the 1970s~\cite{ashkin1970acceleration}, optical trapping and manipulation has evolved into a flexible manipulation technique that benefits multiple disciplines, ranging from biology~\cite{ashkin1987opticalcell,ashkin1987opticalvirus,dao2003mechanics} to quantum physics~\cite{monroe1996schrodinger,kaufman2021quantum}. 
Throughout its development, optical trapping has undergone a fundamental shift from purely electromagnetic field-driven control~\cite{yang2021optical} --  
employing vortex beams~\cite{gahagan1998trapping}, generating counter-intuitive pulling forces~\cite{chen2011optical,dogariu2013optically}, and rotating objects using spin angular~\cite{friese1998optical} and orbital angular momentum~\cite{PhysRevLett.75.826} -- 
 to a more holistic approach in which the material properties of the object are engineered alongside the light field~\cite{spesyvtseva2016trapping}. \\
Indeed, the intrinsic particle characteristics, such as its refractive index and  anisotropy, can be leveraged to directly tune optical forces and torques as demonstrated with birefringent vaterite microparticles~\cite{arita2013laser} and particles with negative relative refractive index contrast ~\cite{prentice2004manipulation, Mao2024switchable}. In parallel, new possibilities have emerged with engineered optical forces on metasurfaces that enable e.g., light-induced steering of meta-vehicles~\cite{andren2021microscopic} or self-stabilized propulsion ~\cite{ilic2019self,michaeli2025direct}. A conceptually different form of control, analogous to atomic systems --  where the optical force flips sign depending on the detuning in respect to the electronic transition -- would offer greater control in force field engineering.\\
To this end, Mie resonances~\cite{ashkin1977observation,stilgoe2008effect} supported by high-permittivity particles~\cite{kuznetsov2016optically} can be leveraged to engineer both the amplitude and the sign of optical forces. In particular, meta-atoms were theoretically predicted to exhibit enhanced and negative polarizabilities, enabling their trapping at intensity minima ~\cite{lepeshov2023levitated,illetschek2026dark,zemanek2002simplified}. Furthermore, anisotropic meta-atoms can be engineered to experience enhanced torques~\cite{toftul2025optical}. Here, we use the term meta-atom, borrowed from metamaterials, for a subwavelength particle whose electric and magnetic multipolar polarisabilities, including their sign, are engineered through its geometry, analogous to the detuning-dependent polarisability of an atom. Along this line, we recently experimentally showed that precise engineering of the coherent superposition of electric and magnetic multipoles supported by a suspended silicon metasurface enables control over the sign of the longitudinal forces it experiences within a standing wave pattern~\cite{afridi2026controlling}.\\
%
Trapping at intensity minima has so far required either a negative refractive-index contrast\cite{prentice2004manipulation,almeida2023trapping}, engineered structured light~\cite{gahagan1996optical,almeida2023trapping,li2025optical}, or active feedback~\cite{dago2024stabilizing}. Recently, Mie-resonant Si nanoparticles were trapped in liquid at the electric-field-free centre of an azimuthally polarised beam~\cite{li2025optical}. while 3D trapping of a solid, high refractive index particle in an intensity minimum was hinted in earlier work~\cite{monteiro2013dynamics}, passive and stable 3D trapping of a solid particle at an intensity minimum of an unstructured field, enabled solely by the particle's own multipolar response, has not been demonstrated.
Beyond its fundamental interest, such trapping regime would greatly benefit  levitation optomechanics~\cite{gonzalez2021levitodynamics}, particularly in precision sensing~\cite{ranjit2016zeptonewton}, ground state cooling  \cite{delic2020cooling, tebbenjohanns2021quantum, magrini2021real, piotrowski2022simultaneous,ranfagni2022two} and mechanical squeezing~\cite{rossi2025quantum,kamba2025quantum}. Indeed, in these applications, mitigating decoherence due to photon recoil~\cite{jain2016direct} is critical for quantum protocols with mesoscopic objects~\cite{roda2024macroscopic,neumeier2024fast,bose2017spin} as is avoiding absorption-induced particle disintegration~\cite{junnemann2025optical}. Recent experimental approaches to achieve this employ active feedback~\cite{dago2024stabilizing} or high refractive index surrounding media~\cite{almeida2023trapping}. In this work, we experimentally demonstrate passive, stable 3D trapping of silicon meta-atoms at moderate vacuum at an intensity minimum, as a way to surpass trapping performance of otherwise predominantly used spherical silica particles.

\begin{figure}
    \centering
    \includegraphics[width=\linewidth]{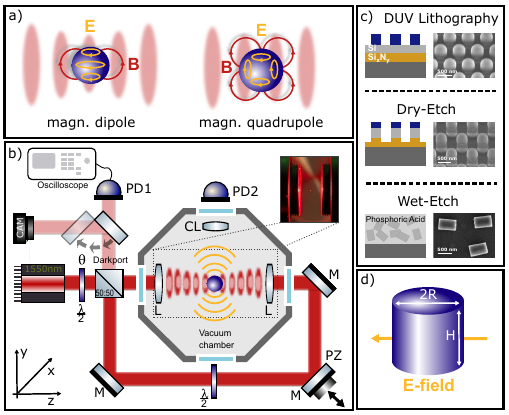}
    \caption{\textbf{Meta-atom trapping with tailored Mie resonances:}    
    \textbf{a)} Illustration of two of the Mie modes supported by the meta-atoms: Magnetic dipole (left) and magnetic quadrupole (right). The magnetic dipole is maximally excited at the magnetic-field antinodes and the magnetic quadrupole at the maxima of the magnetic-field gradient, reflecting the alternating signs in Eq.~\ref{eq:01}.
    \textbf{b)} Experimental setup of the Sagnac interferometer with counterpropagating beams (red) focused by two lenses (L). The particle and its motion are detected on a camera (CAM) and a photodiode (PD1) and an oscillocope. Half waveplates ($\lambda/2$) control the beams's relative power and their polarization while a mirror (M) on a piezo element (PZ) controls the relative phase $\phi$. The collection lens (CL) collects the light scattered (yellow) by the trapped particle (top left inset).
    \textbf{c)} Nanofabrication process flow to fabricate Si particles of controlled sizes with DUV-lithography and etching processes (see SI). 
    \textbf{d)} Cylindrical Si particle of height $H$ and radius $R$. The incident light's electric field component $\mathbf{E}$ is perpendicular to the cylinder's axis.}
    \label{fig:1}
\end{figure}

The emergence of Mie resonances~\cite{mie1908beitrage} is governed by the ratio between the particle size $R$ and the effective wavelength of light within the material $\lambda_\text{eff}$. High-refractive-index materials enable strong resonances within the visible to near-infrared spectrum for particle dimensions of just a few hundred nanometers~\cite{kuznetsov2016optically}. As illustrated in Fig.~\ref{fig:1}a, 
higher-order Mie resonances can coexist and coherently couple with one another. This interference between multipoles does not only modify conventional optical trapping at intensity maxima, resulting in either enhanced or suppressed optical forces~\cite{stilgoe2008effect}, but also enables the counterintuitive trapping of dielectric particles at an intensity minimum~\cite{lepeshov2023levitated,illetschek2026dark,zemanek2002simplified} without requiring specialized beam engineering. This resonant interaction effectively renders the optical force $\mathbf{F}$ tunable, driving a sign change in the particle’s effective polarizability $\alpha_\text{eff}$ that is analogous to the tunability of atomic transitions, yet without involving internal electronic states. Instead of tuning the wavelength $\lambda_\text{eff}$, we here exploit the interplay of individual Mie modes by scanning the particle radius $R$ at a fixed wavelength, inaccessible with atomic resonances. Because Mie modes depend strictly on the size-to-wavelength ratio, scaling the geometry yields the same fundamental force-inversion effect.\\
To describe the tunability of the polarizability $\alpha_\text{eff}$, we assume a standing wave formed by the interference of two linearly polarized, counter-propagating laser beams of equal power $P$. In this specific configuration, the optical gradient force acting on a nanoparticle can be approximated by  $\mathbf{F} \approx \alpha_\text{eff}\nabla I$~\cite{afridi2026controlling}  where $\nabla I$ is the electric field intensity gradient with $I\propto |{\mathbf{E}}^2|$ varying as $|{\mathbf{E}}|^2 \propto\cos^2(kz)$, while $|\mathbf{H}|^2\propto\sin^2(kz)$. The effective particle polarizability $ \alpha_\text{eff}$ is given by 

\begin{equation}\label{eq:01}
    \alpha_\text{eff} \approx [\alpha^{(e)}_{d} - \alpha^{(m)}_{d} ] - [\alpha^{(e)}_{q} -\alpha^{(m)}_{q}] k^2 + [\alpha^{(e)}_{o} - \alpha^{(m)}_{o}] k^4
\end{equation}

where $\alpha_l^{(m)}$ and $\alpha_l^{(e)}$ denote the magnetic and electric polarizabilities for multipole order $l=d, q, o$ (dipole, quadrupole, octupole), proportional to the corresponding Mie scattering coefficients, $\alpha_{l}^{(e)} \propto a_l$ and $\alpha_{l}^{(m)} \propto b_l$ (see SI). The relative minus signs in Eq.~\ref{eq:01} reflect the spatial structure of the standing wave where $E$ and $H$ coupling to their respective multipoles are spatially offset by \(\lambda/4\), with electric-field antinodes coinciding with magnetic-field nodes and vice versa~\cite{afridi2026controlling}. Equation~\ref{eq:01} thus captures how the interplay of these multipoles governs the optical force $\mathbf{F}$ on the particle: a positive effective polarizability ($\alpha_\text{eff} > 0$) drives high-field seeking behavior toward intensity maxima, while a negative one ($\alpha_\text{eff} < 0$) drives low-field seeking behavior toward intensity minima.\\

\textit{Experimental set-up} -- The experimental setup is shown in Fig.~\ref{fig:1}b. A nanoparticle of radius $R$, mass $m$, and refractive index $n_r$ (see inset) is levitated in the focus of two equally linearly polarized, counter-propagating beams, each of power $P  \approx \SI{65}{\milli\watt}$ and vacuum wavelength $\lambda = \SI{1550}{\nano\meter}$, forming a Sagnac interferometer (dark red). The beams inside the vacuum chamber propagate along $z$ and are focused by two lenses (L) with a numerical aperture $\mathrm{NA} = 0.2$ at $z\approx0$. The resulting interference pattern depicted inside the vacuum chamber in Fig.~\ref{fig:1}b forms a series of intensity maxima and minima, where the relative beam phase $\phi$ is controlled by a mirror (M) mounted on a piezoelectric element (PZ). \\
While the scattering force is canceled by the counter-propagating beams, the gradient force generates a series of harmonic trapping potentials along the $z$-axis characterized by the mechanical eigenfrequencies $\Omega_x, \Omega_y, \Omega_z$. 
The mechanical eigenfrequency $\Omega_i$ with $i \in x,y,z$  is set by the corresponding restoring-force gradient $\kappa_i=-\partial F_i/\partial x_i =m \Omega_i^2$ at the equilibrium position, where $F_i$ is the $i$-th Cartesian component of the optical force $\mathbf{F}$. Along $z$, $F_z$ is approximately sinusoidal with amplitude $F_0$, so that $\kappa_z=2kF_0$.
Finally, a half-wave plate ($\lambda/2$) controls the polarization angle $\theta$ of the incident light with respect to the $y$-axis. \\
The Sagnac interferometer enables the detection of the particle motion along $z$ 
at the dark port (light red) by interfering the light scattered by the particle from both interferometer arms on a photodiode (PD1). Due to spurious misalignment we also have access to the radial particle trap frequencies $\Omega_x, \Omega_y$. Simultaneously, a camera images the particle scattering. To highlight the unique characteristics of trapping resonant particles, we compare nanoparticles made of two different materials: standard commercial spherical silica particles (SiO$_2$) and nanofabricated polycrystalline cylindrical silicon particles (Si) of height $H$ and radius $R$, as depicted in Fig.~\ref{fig:1}c-d.\\


\textit{Si particle fabrication} -- While high-quality SiO$_2$ particles are commercially available, well-defined Si particles remain less accessible. Existing production routes typically involve complex synthesis methods~\cite{shi2013monodisperse,sugimoto2020mie} or simpler physical techniques that offer limited control over size dispersion and geometry~\cite{chaabani2019large}. Here, we address this challenge by utilizing a top-down lithography approach, including a sacrificial layer, as illustrated in the process flow shown on the left of Fig.~\ref{fig:1}c. \\
We utilize an ultraflat Si 4" wafer with a thickness of \SI{350}{\micro\meter} (dark gray). A \SI{200}{\nano\meter}-thick Si$_x$N$_\text{y}$ layer (yellow), which serves as a sacrificial layer, is deposited on top of the Si substrate, followed by an polycrystalline Si layer of the target thickness $H$ (light gray). 
The wafer is then spin-coated with a photoresist suitable for top-down deep ultraviolet (DUV) lithography. Next, dry etching is performed using fluorine chemistry in an inductively coupled plasma (ICP) etcher to define cylinders of radius $R$. Following the etch, the photoresist (blue) is stripped using reactive ion etching (oxygen plasma and argon plasma cleaning). The Si cylinders (light gray) are released from the sacrificial layer (yellow) using a wet etch in phosphoric acid 30min at 150$^\circ$C, followed by rinsing in deionized water and sonication in ethanol. Scanning electron microscopy (SEM) images corresponding to these individual steps are shown on the right of Fig.~\ref{fig:1}c.\\

\textit{Enhanced trapping and its characterization} - 
\begin{figure}
    \centering
    \includegraphics[width=\linewidth]{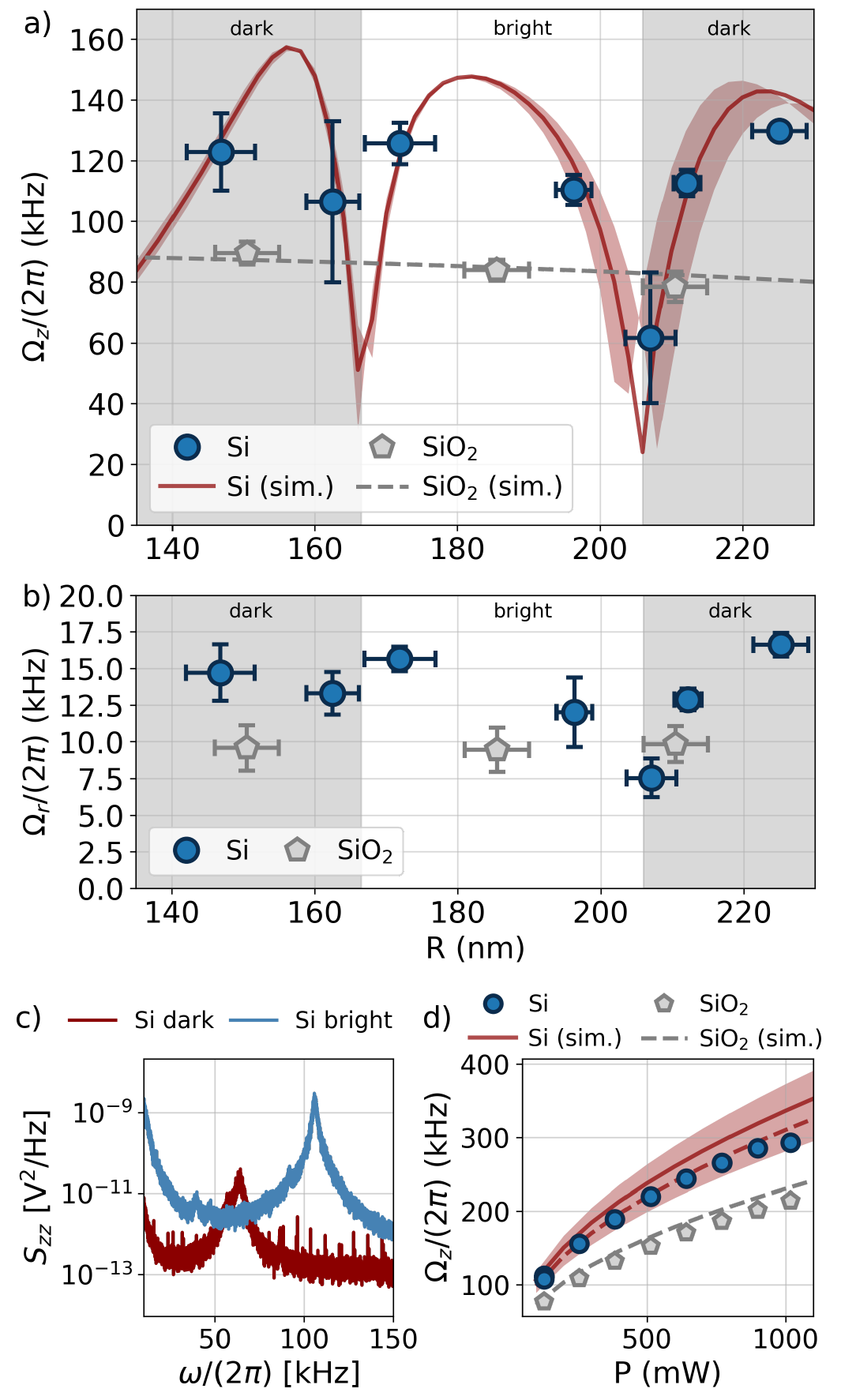}
    \caption{\textbf{Enhanced trapping performance of Si particles}  
    \textbf{a)} Axial trap frequency $\Omega_z$ dependence on $R = R_\text{SEM} + \delta R$ for Si (blue circles) and SiO$_2$ particles (gray pentagons) with $R_\text{SEM}$ being the radius measured in the SEM and $\delta R= \SI{3}{\nano\meter}$ its systematic error. Simulations for Si particles (solid red line) assume $H_\text{SEM}= \SI{605}{\nano\meter}$ with fabrication errors of $\delta H=\pm \SI{10}{\nano\meter}$ (red shaded area), while for SiO$_2$ (dashed gray line) a spherical particle of radius $R$ is assumed. 
    \textbf{b)} Radial trap frequency $\Omega_r$ dependence on $R = R_\text{SEM} + \delta R$ for Si (blue circles) and SiO$_2$ particles (gray pentagons). The gray shaded areas highlight radii where particles are trapped at the intensity minimum ($z= \pm \lambda/4$)  due to $\alpha_\text{eff}<0$. Data points and vertical errorbars are the mean and STD of five (three) independent trapping measurements for Si (SiO$_2$). Horizontal errorbars represent the STD of the SEM size measurement (see SI).
   \textbf{c)} Displacement PSD of Si particles of $R_\text{SEM}=(204.1 \pm 3.5)\:\SI{}{\nano\meter}$ , $H_\text{SEM}= (609.6 \pm 12.2)\:\SI{}{\nano\meter} $ (red) and $R_\text{SEM}=(193.3 \pm 2.5)\:\SI{}{\nano\meter}, H_\text{SEM}= (594.2 \pm 6.4)\:\SI{}{\nano\meter}$ (blue) exhibiting a Lorentzian shaped response function with a well-defined $\Omega_z$.
    \textbf{d)} Power dependence of $\Omega_z$ for Si ($R_\text{SEM}=(209.2 \pm 1.9)\:\SI{}{\nano\meter},  H_\text{SEM}= (605 \pm 1.9)\:\SI{}{\nano\meter}$, blue circles) and SiO$_2$ ($R=\SI{207.5}{\nano\meter}$, gray pentagons) following $\Omega_{z}\propto\sqrt{P}$ as expected from theory. Simulations assuming a nominal height that takes the systematic error of the SEM into account
    ($R = R_\textbf{SEM} + 3\text{nm} =\SI{212}{\nano\meter}, H=H_\text{SEM} +6\text{nm}=\SI{611}{\nano\meter}$, solid red line) overestimates $\Omega_z$ in comparison to the best match  ($R=\SI{210}{\nano\meter}, H=\SI{613}{\nano\meter}$, dashed red line). The shaded area assumes $\delta R=\pm \SI{2}{\nano\meter},\delta H=\pm \SI{2}{\nano\meter}$ from the nominal size.
   }
    \label{fig:2}
\end{figure}

Figure~\ref{fig:2} displays how the trap frequencies $\Omega_i$ related to the optical restoring-force gradient depend on the radius $R = R_\text{SEM} + \delta R$ of individual silicon cylinders at a pressure $p=\SI{5}{\milli \bar}$, where we assume a systematic error of the radius measured in the SEM $R_\mathrm{SEM}$ by $\delta = \SI{3}{\nano\meter}$.
Our primary focus lies on the axial trap frequency $\Omega_z=\sqrt{\kappa_z/m}$ 
which experiences the largest intensity gradient, leading to the highest $\Omega_z$. Note that we ensure trapping close to the focus of the two beams where the trap frequency is maximum, allowing for direct comparison across different particle sizes.  The mechanical eigenfrequency $\Omega_z$ is extracted by applying a Lorentzian fit to the power spectral density (PSD) of the displacement of a thermally driven harmonic oscillator (see SI). For comparison with the non resonant case, we also repeated the experiment with SiO$_2$ particles. 
As can be seen in Fig.~\ref{fig:2}a, Si particles with varying radii within $\SI{145}{\nano\meter}\leq R \leq \SI{225}{\nano\meter}$ and height $H= (604.4\pm 11.2)\SI{}{\nano\meter}$ 
exhibit trap frequencies (blue circles) spanning $\Omega_z/(2\pi) \approx [\SI{60}{\kilo\hertz},\SI{125}{\kilo\hertz}]$. This behavior stands in strong contrast to spherical SiO$_2$ particles (gray pentagons) that reach for the same trapping conditions nearly constant $\Omega_z/(2\pi) \approx [\SI{80}{\kilo\hertz},\SI{85}{\kilo\hertz}]$. The data points and errorbars represent the mean and standard deviation (STD) obtained from five  (three) individual measurements of the same Si (SiO$_2$) particle batch. \\
The experimental data are compared with COMSOL Multiphysics simulations utilizing a scattered-field formulation. The system models a single nanoparticle centered within a spherical domain bounded by perfectly matched layers (PML) to suppress spurious reflections, with an ideal standing wave serving as the background field. Optical forces are computed via the Maxwell stress tensor method~\cite{jackson2021classical}. The refractive indices at the trapping wavelength are taken as $n_{\text{Si}} = 3.48 + \text{i} 5.3 \times 10^{-11}$ for Si~\cite{degallaix2013bulk} and $n_{\text{SiO}_2} = 1.46 + \text{i} 5 \times 10^{-9}$ for $\text{SiO}_2$~\cite{palik1998handbook}. At the experimental pressure of $p = \SI{5}{\milli\bar}$, laser-induced particle heating~\cite{junnemann2025optical} and induced thermal gradients are expected to have a minimum influence on the conservative optical forces. The $\text{SiO}_2$ particles are modeled as spheres of radius $R$, whereas the Si particles are modeled as cylinders of radius $R$ and height $H$. Notably, the cylinders are assumed to be aligned with their cylinder axis parallel to the beam propagation axis ($z$-axis, see Fig.~\ref{fig:1}d). This orientation yields good agreement between the simulations and the experimental data of Figs.~\ref{fig:2}. 
Assessing the cylinder's rotational stability through optical torque calculations is beyond the scope of this work. \\
As highlighted by the simulations for Si particles (solid red line), the dependence of $\Omega_z$ on $R$ is highly non-trivial, displaying several resonances separated by two anti-resonant features at $R \approx \SI{165}{\nano\meter}$ and $\SI{205}{\nano\meter}$. Conversely, for $\text{SiO}_2$ particles (gray dashed line), the simulations predict a nearly constant $\Omega_z$ across all radii, with a monotonic trend that closely matches the experimental data. A similar trend is observed for the radial trap frequencies $\Omega_r$ depicted in Fig.~\ref{fig:2}b but less pronounced than for $\Omega_z$ due to the inherently lower absolute values of the radial trap frequencies.  Since $\Omega_r$ depends sensitively on the transverse beam profiles and on the residual misalignment of the two beams, we restrict the quantitative comparison with simulations to the axial direction. Overall, comparing Si and $\text{SiO}_2$ particles reveals a significant enhancement in trap stiffness, and consequently a gain in optical forces, quantified by the ratio $\eta = \Omega_z^\text{Si}/\Omega_z^{\text{SiO}_2} 
\approx 45\%$. Theoretically, this enhancement can reach $\eta \approx 100\%$ at an optimized radius of $R \approx \SI{150}{\nano\meter}$.\\
Figure~\ref{fig:2}c displays example PSDs of the particle displacement 

\begin{equation}
S_{zz}(\omega) = \frac{2k_BT\Gamma/m}{(\Omega_z^2-\omega^2)^2+\Gamma^2\omega^2} 
\end{equation} along $z$ for Si particle of radii $R_\mathrm{SEM} = (204.1 \pm 3.5)\SI{}{\nano\meter}$ (red) trapped at the intensity minimum (dark) and $R_\mathrm{SEM} = (193.3 \pm 2.5)\SI{}{\nano\meter}$ (blue) trapped at the intensity maximum (bright), exhibiting different $\Omega_z$ under identical trapping conditions. For both dark and bright trapping we recover the expected Lorentzian shape of a driven, underdamped harmonic oscillator. \\ 
In Fig.~\ref{fig:2}d we confirm the expected power scaling $\Omega_z\propto\sqrt{P}$ for both a Si particle of $R_\mathrm{SEM} = (209.2\pm 1.9)\SI{}{\nano\meter}$ being trapped in the dark (blue circles) and a SiO$_2$ particle of $R = \SI{207.5}{\nano\meter}$ (gray pentagons), where the Si particle exhibits significantly higher trapping frequencies with a maximum of $\Omega_z/(2\pi)\approx \SI{290}{\kilo\hertz}$. 
The simulation for Si particles (red shaded area) accounts for fabrication errors of $\delta H\approx \pm\SI{2}{\nano\meter}$ and $\delta R \approx\pm\SI{2}{\nano\meter}$ where the nominal size of the cylinder ($R= \SI{212}{\nano\meter}, H=\SI{611}{\nano\meter}$, red solid line) surpasses the best match (red dashed line) assuming $\delta R \approx-\SI{2}{\nano\meter},\delta H \approx +\SI{2}{\nano\meter}$. For SiO$_2$, the simulation also slightly overestimates the experimental result which we attribute to an underestimation of the experimental power $P$. \\
The availability of such a rich force landscape for Si particles arises because already the smallest particle considered ($R = \SI{150}{\nano\meter} \approx \lambda/10$) approaches the limit of the dipole approximation. 
Consequently, higher-order Mie modes play a significant role across the entire size range investigated here (see SI) where individual multipolar Mie coefficients $a_l,b_l$ (or generalized Mie coefficients for non-spherical geometries) exhibit a non-trivial dependence on the particle radius $R$~\cite{lepeshov2023levitated}, meaning that their respective phase and amplitude contributions can sum constructively or partially cancel. 
Specifically, for the pronounced resonance observed at $R \approx \SI{150}{\nano\meter}$, 
the contributions from the magnetic dipole ($b_1$) are only partially compensated by the electric dipole ($a_1$), leading to an enhanced effective polarizability $\alpha_{\text{eff}}$ and therefore a significantly larger restoring force. Conversely, at the anti-resonance around $R \approx \SI{205}{\nano\meter}$, the multipolar contributions undergo destructive interference, where the phase shift between the dominant multipoles suppresses the overall scattering amplitude and minimizes the net optical gradient force. \\
%
%
More remarkably, the interplay of multipole polarizabilities (see Eq.~\ref{eq:01}) also enables 
 negative effective polarizabilities $\alpha_\text{eff}<0$ attracting the Si particle toward an intensity minimum, 
 in contrast to SiO$_2$ particles always trapped at the intensity maximum. The different trapping scenarios suggested by simulations are highlighted in Fig.~\ref{fig:2}a as dark trapping at intensity minima $|z|=\lambda/4$ (gray shaded regions) and bright trapping at intensity maxima $z=0$ (white regions).\\ 

\begin{figure*}
    \centering
    \includegraphics[width=1\linewidth]{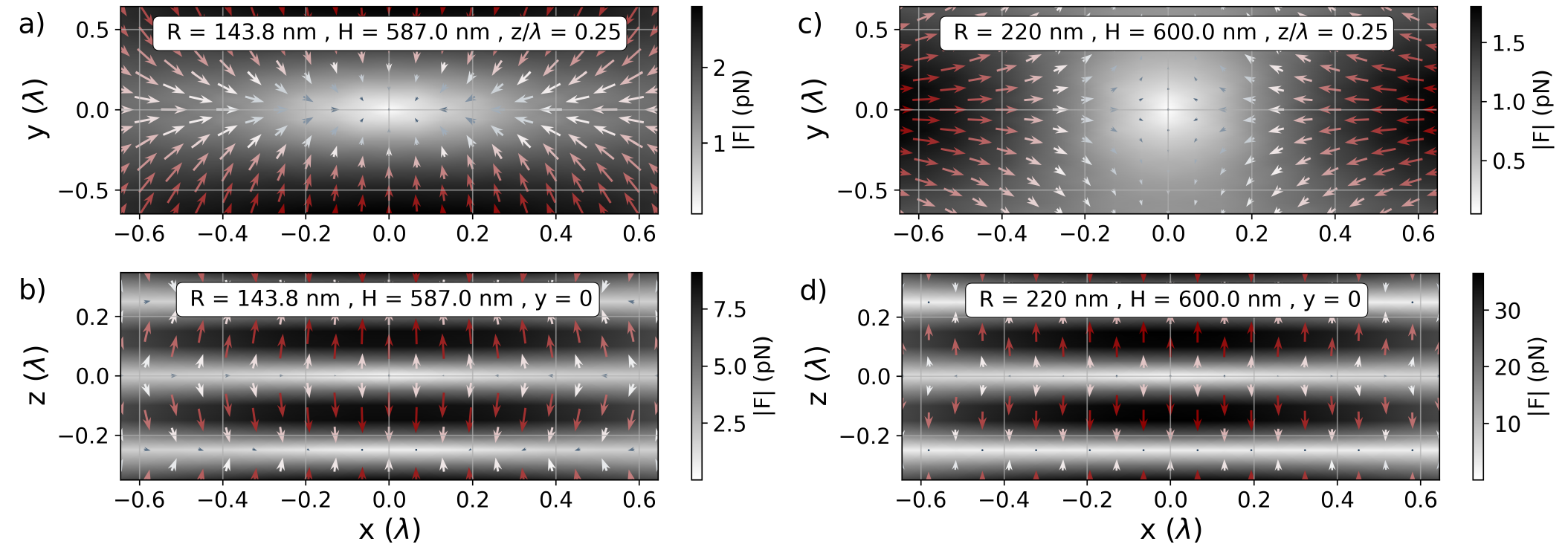}
    \caption{\textbf{Force field simulations experienced by Si meta-atoms} being trapped at the intensity minimum. Force field $\mathbf{F}$ (arrows) and force magnitude $|\mathbf{F}|$ (shaded background) for \textbf{a-b)} $R=\SI{143.8}{\nano\meter}, H=\SI{587}{\nano\meter}$ and \textbf{c-d)}  $R=\SI{220}{\nano\meter}, H=\SI{600}{\nano\meter}$.  \textbf{a)} The particle is trapped at $x,y=0$ and  \textbf{b)} $z=\pm\lambda/4$.  \textbf{c)} The particle is repelled from the center along $y$, trapped at $x=0$, and \textbf{d)} $z=\pm\lambda/4$. 
    }
    \label{fig:03}
\end{figure*}

To gain further insight into the dark trapping regime, we simulate the optical force map $\mathbf{F}$ acting on a Si cylinder 
in both the $xy$- and $zy$-planes. We compare the two specific particle sizes ($R = \SI{143.8}{\nano\meter}, H=\SI{587}{\nano\meter}$ and $R  = \SI{220}{\nano\meter}, H=\SI{600}{\nano\meter}$) associated with dark trapping in Fig.~\ref{fig:2}a. 
The two dark-trapping regimes differ in the origin of their transverse stability. Figures~\ref{fig:03}a and \ref{fig:03}b show the vector force field $\mathbf{F}$ (arrows), which creates a potential minimum at $x = y = 0$ and $z = \pm\lambda/4$ (indicated by the white background area) for $R=\SI{143.8}{\nano\meter}$. In contrast, for $R=\SI{220}{\nano\meter}$, while the axial force component still creates a local potential minimum at $z = \pm\lambda/4$ (see Fig.~\ref{fig:03}d), the transverse force along the $y$-axis (see Fig.~\ref{fig:03}c) expels the particle from the beam axis in the aligned configuration (see Fig.~\ref{fig:03}c), rendering the trap unstable. Experimentally, however, we observe stable 3D confinement in this size regime. We attribute this to a small transverse offset between the two parallel, counterpropagating beams, with their axes at \(y=\pm\delta y\). Between the beam axes, outward transverse force contributions associated with the two beams act in opposite directions, providing stable confinement along \(y\).
Although the absolute radial trapping position could not be resolved experimentally, we expect this displacement to be small because we observe the highest trap frequencies for $R\approx \SI{220}{\nano\meter}$ (see Fig.~\ref{fig:2}a).
Note the difference between the two particle regimes. For $R=\SI{143.8}{\nano\meter}$, the particle response leads to stable 3D trapping, while for $R=\SI{220}{\nano\meter}$ an additional transverse offset between the beams is required for stable 3D confinement.\\

\textit{Scattering patterns and trapping in the dark} -
\begin{figure}
    \centering
    \includegraphics[width=\linewidth]{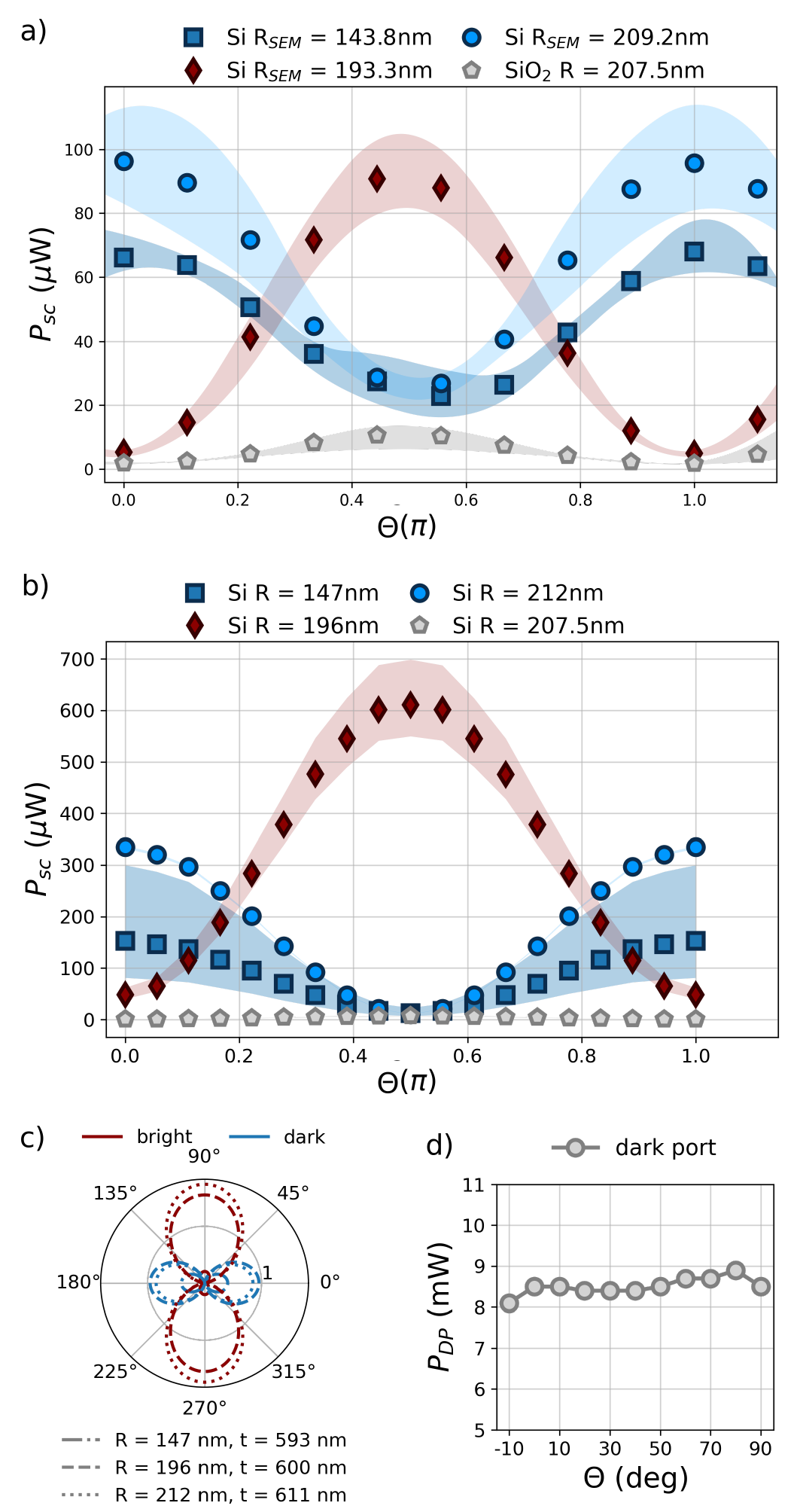}
    \caption{\textbf{Scattering pattern reveals trap position.} 
    \textbf{a)} Scattered power $P_{sc}$ being collected (CL) along $y$ for varying polarization angle $\theta$. Si particles of $R_\text{SEM}=\SI{143.8}{\nano\meter}, H_\text{SEM}=\SI{587.2}{\nano\meter}$ (blue squares) and $R_\text{SEM}=\SI{209.2}{\nano\meter}, H_\text{SEM}=\SI{605}{\nano\meter}$ (blue circles) exhibit a maximum scattered power at $\theta=n\pi$ with $n\in\mathbb{Z}$ corresponding to the polarization axis of $\mathbf{E}$. 
    In contrast Si particles of $R_\text{SEM}=\SI{193.3}{\nano\meter}, H_\text{SEM}=\SI{594.2}{\nano\meter}$ (red diamonds) scatter maximum power along $y$ at $\theta=n \:\pi/2$ perpendicular to $\mathbf{E}$ same as SiO$_2$ (gray pentagons). 
    The shaded area is the respective minimum and maximum value of three measurements with individual particles. 
    \textbf{b)} Simulation of $P_{sc}$ collected with CL. Si particles scatter maximally at $\theta = n\pi$ parallel to $\mathbf{E}$ (
    $R= \SI{147}{\nano\meter}\pm\SI{5}{\nano\meter}, H= \SI{593}{\nano\meter}\pm \SI{6}{\nano\meter}$ (blue dot dashed)  and 
    $R= \SI{212}{\nano\meter}\pm\SI{2}{\nano\meter}, H= \SI{611}{\nano\meter}\pm \SI{2}{\nano\meter}$ (blue dashed))  or at $\theta= n\pi/2$ ($R= \SI{196}{\nano\meter}\pm\SI{3}{\nano\meter}, H= \SI{600}{\nano\meter}\pm \SI{6}{\nano\meter}$ (red solid)) perpendicular to $\mathbf{E}$ same as SiO$_2$ ($R=\SI{207.5}{\nano\meter}$ (gray dotted)).
    \textbf{c)} Simulation of the scattering pattern in the $xy$-plane in polar coordinates of Si particles of various sizes  being held at the intensity maximum (red) and intensity minimum (blue). The scattering pattern at different positions are rotated by $\pi/2$ for all particle sizes. 
    \textbf{d)} Control measurement of the dark port power $P_\text{DP} = \SI{8.5}{\milli\watt} \pm \SI{0.3}{\milli\watt}$  being independent of the polarization angle $\theta$. 
    }
    \label{fig:04}
\end{figure}
While we have thus far claimed that trapping can occur at an intensity minimum for certain radii of Si meta-atoms, this assertion has only been supported by simulations (see Fig~\ref{fig:2} and Fig~\ref{fig:03}). Experimentally, the enhanced trapping frequencies (see Fig.~\ref{fig:2}) alone are insufficient to determine the trapping position $z$. To experimentally support trapping at the intensity minimum, we exploit the dependence of the directionality of the scattering-pattern on the dominating multipole contribution. The main contributions to the scattering pattern stem from the electric $\mathbf{p} = \alpha_d^{(e)} \mathbf{E}$ and  magnetic dipole moments $\mathbf{m} = \alpha_d^{(m)}\mathbf{H}$. For a dipole scatterer, it is well known that no power is scattered to the far field along the dipole axis, while maximum power is detected perpendicular to the dipole axis. Due to $\mathbf{E} \perp \mathbf{B}$, the scattering directions for $\mathbf{p}$ and $\mathbf{m}$ are perpendicular to each other. In combination with $|\mathbf{E}|\neq 0, |\mathbf{H}|=0$ ($|\mathbf{E}| = 0, |\mathbf{H}|\neq 0$)  at the intensity maximum $z=0$ (minimum $z=\pm\lambda/4$), the scattering pattern predominantly contains contributions from the electric dipole moment $\mathbf{p}$ at $z=0$, while at $z=\pm\lambda/4$ only the magnetic dipole moment $\mathbf{m}$ contributes at the dipole level. Higher-order multipoles modify the scattering pattern quantitatively and are fully accounted for in the simulations of Figs.~\ref{fig:04}b,c. Hence, the directionality of the scattering provides us a direct measurement of the position of the particles $z$ within the standing wave pattern.\\
The Rayleigh-scattered light (yellow) is collected using a lens (CL) and detected by a photodiode (PD2) positioned along the $y$-axis (see Fig.~\ref{fig:1}b). The polarization angle $\theta$ is defined with respect to the $y$-axis, such that $\theta = 0$ and $\theta = \pi/2$ correspond to $y$-polarized and $x$-polarized light, respectively.
For a $\text{SiO}_2$ particle operating near the dipolar regime ($R = \SI{207.5}{\nano\meter} \ll \lambda$) trapped in a $x$-polarized light field at $z = 0$, the maximum Rayleigh scattering occurs along the $y$-axis (PD2). Conversely, for $y$-polarized light, the collected power reaches a minimum (gray pentagons), as shown in Fig.~\ref{fig:04}a. For high-refractive-index particles, however, this behavior changes substantially depending on the stable trapping position, corresponding to the effective polarizability $\alpha_{\text{eff}}$ (see SI). \\
In Fig.~\ref{fig:04}a, we investigate three different $\text{Si}$ particle sizes from Fig.~\ref{fig:2}a, exhibiting different responses as the polarization angle $\theta$ is varied. Maximum scattering perpendicular to the polarization axis ($\theta = \pi/2$), mirroring the behavior of the $\text{SiO}_2$ particle, coincides with trapping at the intensity maximum ($z = 0$) for $R_\mathrm{SEM} = \SI{193.3}{\nano\meter}$ (red diamonds). In contrast, maximum scattering parallel to the polarization axis ($\theta = 0$) signatures trapping at the intensity minimum ($z = \pm\lambda/4$) for both $R_\mathrm{SEM} = \SI{143.8}{\nano\meter}$ (blue squares) and $R_\mathrm{SEM} = \SI{209.2}{\nano\meter}$ (blue circles). \\
To unambiguously rule out optical misalignment artifacts, we continuously rotate the incident polarization angle from $\theta= 0$ to $\theta= \pi$ while simultaneously measuring the scattered power along $y$. For each trapping regime, we recover the expected sinusoidal dependence on $\theta$. Notably, the absolute scattered power is significantly lower for low-refractive index material. The data points represent the mean values and the shaded regions the maximum and minimum spread calculated from three independent measurements.\\
For comparison, we simulate the angular dependence of the collected scattered power using \textsc{Comsol}, as shown in Fig.~\ref{fig:04}b. These simulations confirm the observed experimental trends for particle radii of $R = \SI{147}{\nano\meter}$, $\SI{196}{\nano\meter}$, and $\SI{212}{\nano\meter}$, which incorporate a systematic SEM calibration offset of $\delta R= \SI{3}{\nano\meter}$. The corresponding angular scattering patterns~\cite{tebbenjohanns2019optimal,maurer2022} are illustrated in Fig.~\ref{fig:04}c for Si particles located at either the intensity maximum (red) or the intensity minimum (blue). Figure~\ref{fig:04}c clearly demonstrates a $\pi/2$ rotation of the far-field scattering patterns between the intensity maxima ($z = 0$) and intensity minima ($z = \pm\lambda/4$). This rotation is independent of the particle size due to position dependent zero crossings of $\mathbf{E}$ and $\mathbf{B}$, with the lobe amplitude directly representing the relative scattered power in each configuration.\\
To exclude the possibility that the observed intensity variations arise from misalignment or drifts during the rotation of the beam polarization,  we monitor the power at the dark port $P_\text{DP}$ (see Fig.~\ref{fig:1}b). As shown in  Fig.~\ref{fig:04}d, the dark-port power yields $P_\text{DP} = \SI{8.5}{\milli\watt} \pm \SI{0.3}{\milli\watt}$ over the full range $\theta \in [0,\pi/2]$. 
The excellent match between experiment and theory allows us to unambiguously confirm dark trapping  for $R_\mathrm{SEM}=\SI{143.8}{\nano\meter}$ and $\SI{209.2}{\nano\meter}$. Finally, we emphasize that trapping at an intensity minimum does not imply the absence of light at the particle position. As discussed in~\cite{lepeshov2023levitated}, high-refractive index particle can significantly modify the local intensity distribution through their strong polarizability, and the scattering rate is not necessarily suppressed and can, in fact even be enhanced.


\textit{Discussion} -
The experimental demonstration of low- and high-field-seeking behavior depending on $R/\lambda_\text{eff}$ is fully analogous to atoms where the detuning in respect to electronic transitions governs the attractive or repulsive nature of the optical force. 
In contrast to atoms though, here we identified with $R$ an additional continuously tunable parameter, 
while atoms are bound by their constituents. Furthermore, low- and high-field seeking behavior enables trapping in close proximity to surfaces with controllable subwavelength distances~\cite{vetsch2010optical}
, with the distance being independent from the wavelength like in retro-reflected traps~ \cite{diehl2018optical} enabling studies of surface forces~\cite{casimir1948attraction, margenau1939van} or coupling to nanophotonic devices~\cite{magrini2018near}. \\
The complex trend of $\Omega_z$ and photon recoil rate $\Gamma$  with $R$ as discussed in~\cite{lepeshov2023levitated} now allows us to reduce the photon recoil heating rate when it becomes the limiting decoherence rate. Alternatively, it also enables us to reach a regime where high-photon scattering rates facilitate ground-state cooling via active feedback, as higher levels of gas damping can be tolerated while remaining shot-noise limited. Overall, the ability to tune particle confinement and scattering rates~\cite{lepeshov2023levitated} allows us to maximize the $Q \times f$ product, a central figure of merit for implementing quantum protocols in optomechanical systems.\\
Despite these promising trapping characteristics, we currently observe particle loss at moderate background pressures ($~\geq \SI{0.1}{\milli\bar}$)~\cite{junnemann2025optical}. Preliminary indications suggest that this loss mechanism may be mitigated by active feedback cooling~\cite{monteiro2013dynamics}, but absorption-induced heating is expected to elevate the particle’s internal temperature~\cite{junnemann2025optical}, raising the need to consider less absorbing materials to reach the quantum regime. 

\textit{Conclusions} -
Leveraging reliable top-down nanofabrication of silicon cylinders with characteristic dimensions well below \SI{1}{\micro\meter}, we demonstrated robust, passive 3D trapping at a intensity minimum. Furthermore, we achieved substantially enhanced trapping performance, characterized by markedly increased absolute trap frequencies. These advanced capabilities remain fundamentally inaccessible to conventional low-refractive-index dipole scatterers widely used in levitated optomechanics. 
This work opens up promising new avenues for optical force engineering and precise optomechanical control in both gaseous and liquid environments. In the context of levitation optomechanics, it represents  a step toward levitated systems that enable creating inverted potentials without requiring beam engineering. Such potentials were proposed as a route to macroscopic quantum superpositions ~\cite{roda2024macroscopic,neumeier2024fast} and are already the subject of extensive experimental effort ~\cite{tomassi2026accelerated,duchavn2025nanomechanical,dago2024stabilizing}.


\vspace{0.5cm}
\textbf{Acknowledgements:} This research was supported by the European Research Council (ERC) through grant Q-Xtreme ERC 2020-SyG (grant agreement number 951234). We acknowledge valuable discussions with the Q-Xtreme synergy consortium. \\

\textbf{Author contributions} - 
BL fabricated the device, performed the measurements and analyzed the data.   
BL and AA performed the numerical simulations,
NM developed the theoretical expressions. NM and RQ conceptualized the experiments. All authors discussed the results and contributed to writing the manuscript.\\

\textbf{Competing Interests} - 
The authors declare no competing interests.\\

\textbf{Data availability} - 
The data supporting the findings of this study are available within the
article and its Supplementary Information. The source data files are
available via the ETH Zürich Research Collection at TBA.



%

\end{document}